\documentclass[prb, twocolumn,showpacs]{revtex4}
\begin{document}

\title{Linear Chains of Styrene and Methyl-Styrene
Molecules and their Heterojunctions on Silicon:  Theory
and Experiment}

\author{George Kirczenow}

\affiliation{Department of Physics, Simon Fraser
University, Burnaby, British Columbia, Canada V5A 1S6}

\author{Paul G. Piva}

\affiliation{National Institute for Nanotechnology,
National Research Council of Canada, Edmonton, Alberta
T6G 2V4, Canada, and Department of Physics, University of
Alberta, Edmonton, Alberta T6G 2J1, Canada}

\author{Robert A. Wolkow}

\affiliation{National Institute for Nanotechnology,
National Research Council of Canada, Edmonton, Alberta
T6G 2V4, Canada, and Department of Physics, University of
Alberta, Edmonton, Alberta T6G 2J1, Canada}

\date{\today}


\begin{abstract}

We report on the synthesis, STM imaging and theoretical
studies of the structure, electronic structure and
transport properties of linear chains of styrene and
methyl-styrene molecules and their heterojunctions on
hydrogen-terminated dimerized silicon (001) surfaces. The
theory presented here accounts for the essential features
of the experimental STM data including the nature of the
corrugation observed along the molecular chains and the
pronounced changes in the contrast between the styrene
and methyl-styrene parts of the molecular chains that are
observed as the applied bias is varied. The observed
evolution with applied bias of the STM profiles near the
ends of the molecular chains is also explained.
Calculations are also presented of electron transport
along styrene linear chains adsorbed on the silicon
surface at energies in the vicinity of the molecular HOMO
and LUMO levels.  For short styrene chains this lateral
transport is found to be due primarily to direct electron
transmission from molecule to molecule rather than
through the silicon substrate, especially in the
molecular LUMO band. Differences  between the calculated
position-dependences of the STM current around a junction
of styrene and methyl-styrene molecular chains under
positive and negative tip bias are related to the nature
of lateral electron transmission along the molecular
chains and to the formation in the LUMO band of an
electronic state localized around the heterojunction.

\end{abstract}

\maketitle

\section{Introduction}

During the last few years there has been growing interest
in molecular electronics, stimulated largely by the
experimental realization of molecular
wires,\cite{Bumm,Reed1,Datta} systems in which a single
organic molecule or a few molecules carry an electric
current between a  pair of metal contacts. In some cases
such systems exhibit switching behavior and/or  negative
differential resistance,\cite{Reedsw,Collier,Donhauser}
phenomena that may be exploited in future molecular
electronic devices. Hybrid molecular/semiconductor
nano-electronic devices are another intriguing
possibility and fundamental research that may ultimately
lead to their creation is also  being pursued at the
present
time.\cite{Wolkow,Lopinski,Cho,Hofer,Hersam,Tong,Rakshit,Wang} 
Recently, techniques have been developed that make it
possible to grow an oriented linear chain of styrene
molecules on a hydrogen-terminated silicon substrate
beginning at a predefined point on the
substrate.\cite{Lopinski,Tong} The electronic structure
and STM images of these homomolecular chains of styrene
have also been modelled theoretically.\cite{Cho,Hofer}
However, no studies (experimental or theoretical) of
chains of styrene molecules with chemical substituents 
attached have been reported to date. In the work reported
in the present article we extend the experiments and
theory to a related system, a {\em heteromolecular} chain
of molecules on a silicon substrate where the chemical
identity of the molecular species changes abruptly across
a heterojunction from styrene to methyl-styrene. As the
chemical composition across the molecular chain varies
solely by the presence (or absence) of a substituent (in
this case a methyl group) on the aromatic ring,  the
$\pi$-$\pi$* stacking that is an essential prerequisite
for wire-like conduction along the chain remains
undisrupted.

Some intriguing questions regarding the self-assembled
lines of molecules on silicon are: To what extent do the
molecules that form the line communicate with each other
electronically? Is electrical conduction along the line
possible, and if so is such conduction dominated by
transport through the molecules themselves or through the
silicon substrate? Calculations of lateral transport
along the molecular chain can in principle answer these
questions. However, to date such calculations have not
been reported for molecular chains on silicon, although
molecular band formation in the styrene chains due to
orbital overlap between molecules has been
studied,\cite{Cho,Hofer} as has conduction through stacks
of aromatic biphenyldithiol molecules not attached to a
substrate.\cite{Rochefort} The role of molecule-molecule
interactions in vertical conduction through molecules
between metal contacts has also been investigated
theoretically.\cite{Yaliraki,Magoga,Lang,Kushmerick1,Liu1}
 
To date it has not been feasible experimentally to
connect a pair of probes to the two ends of a chain of
molecules on silicon so as to measure the lateral
transport directly: STM studies of the molecular chains
measure  vertical transport between the silicon substrate
and a single metal tip via the adsorbed
molecules.\cite{Lopinski,Tong} However, vertical
transport measurements may yield indirect experimental
information relevant to lateral transport, particularly
if the molecular chain is inhomogeneous. A heterojunction
between chains of similar but distinct molecules such as
styrene and methyl-styrene on silicon or the end of a
line of molecules may be a suitable inhomogeneity for
such studies, but theoretical modeling of such systems is
clearly needed in order to extract meaningful information
of this kind from experimental data. 

A realistic model is presented here of the structure,
electronic structure and electronic transport in chains
of styrene and methyl-styrene molecules on
hydrogen-terminated  dimerized silicon (001) surfaces.
The model accounts for the main properties of
heterojunctions of styrene and methyl-styrene molecular
chains on silicon  that have been observed in our STM
experiments, as well as the experimentally observed
phenomena associated with the ends the molecular lines.
The first calculations are  presented of lateral
transport though short styrene chains adsorbed on the
silicon surface. It is found that lateral electron
transport along an 8-molecule styrene chain at energies
corresponding to the molecular HOMO and LUMO bands is
primarily due to direct electron transmission from
molecule to molecule rather than transmission via the
silicon substrate. This dominance of molecule-to-molecule
transmission over transmission through the substrate is
stronger for the LUMO band than the HOMO band. This
difference is related to pronounced differences between
the calculated position-dependences of the STM (vertical
transport) current around the junction of styrene and
methyl-styrene molecular chains under positive and
negative tip bias.

This article is organized as follows: In Section
\ref{Expt} we present experimental results detailing the
growth and STM imaging characteristics of heteromolecular
chains of styrene and methyl-styrene on silicon.  In
Sections \ref{structure} and \ref{electronicstruct} we
describe our modeling of the structure and electronic
structure of the molecular chains.  The approach used in
our transport calculations is summarized in Section
\ref{transport}. Our theoretical results for vertical
transport  in the STM geometry in the vicinity of a
heterojunction of molecular lines are presented and
compared with experiment in Section \ref{vertical}. Our
theoretical results   regarding lateral transport along
the molecular chains and how this may influence STM images
are presented in  Section \ref{lateral}. Our experimental
and theoretical findings regarding the structure and STM
images of the ends of the molecular chains are discussed
in Section \ref{ends}.  Finally, our conclusions are
summarized in Section
\ref{conclusions}.

\section{Experiment}
\label{Expt}

The experiments were performed on a hydrogen terminated
Si(100)-2$\times$1 surface. Samples were cleaved from an
arsenic-doped (0.005 $\Omega\cdot$cm) Si (100) oriented
wafer, mounted in molybdenum sample holders, and
load-locked into an ultrahigh vacuum (UHV) chamber
(background system pressure $<$ 1$\times$$10^{-10}$ Torr).
Samples were degassed for 8 hours at 700$^{\circ}$C. 
Removal of the surface oxide, and
sublimation/replanarisation of the surface layers was
achieved by repeated thermal cycling to 1250$^{\circ}$C. 
Sample temperatures were monitored using a calibrated
infrared pyrometer.  During the annealing procedure,
sample heating was discontinued if at any point the
monitored system pressure exceeded 4$\times 10^{-10}$
Torr.  This sample preparation method produces 100
oriented silicon surfaces with well ordered silicon
terraces (2$\times$1 surface reconstruction), and a
surface defect concentration below 5\%.

The vacuum preparation chamber was then filled with
hydrogen gas (1$\times$$10^{-6}$ Torr), and a hot tungsten
filament (T $\sim$ 1600$^{\circ}$C) situated 10 cm from
the sample was used to provide a flux of reactive atomic
hydrogen to the surface. The sample was heated to
330$^{\circ}$C and exposed to the atomic hydrogen flux for
13 minutes.  These conditions allowed the formation of a
quasi-saturated monolayer of silicon monohydride with a
2$\times$1 surface reconstruction.\cite{Boland}

One-dimensional molecular organic heterowires were grown
along dimer rows of the as-prepared H:Si (100) surfaces
using a vacuum phase reaction  studied earlier by Lopinksi
and co-workers.\cite{Lopinski} In the case of the styrene
reaction with the H:Si surface, the double carbon bond in
the vinyl group at the base of styrene molecule reacts
with a surface silicon radical (corresponding to a
hydrogen vacancy on the H:Si surface) to produce a
silicon-carbon bond and a carbon radical (centred on the
carbon atom bound to the aromatic ring).  The carbon
centred radical  then abstracts a hydrogen atom from an
adjacent dimer on the same row (and from the same side of
the dimer - never from the other (diagonal) atom),
producing a newly reactive site.   Successive reactions
lead to the well ordered linear structures reported in Ref.
\onlinecite{Lopinski}.  Several substituted styrene
compounds have been studied in our group and have
demonstrated this self-directed line growth mechanism on
H:silicon.\cite{submitted} Here two such compounds are
used,  styrene and 4-methyl-styrene, to fabricate
one-dimensional organic heterowires.
  
On the H:Si(100) surface, residual silicon radicals
typically exist at the level of a few percent and are used
in this work as initiation sites for the self-directed
growth of 4-methyl-styrene/styrene heterowires. The
molecular heterowires were fabricated by sequentially
introducing 4-methyl-styrene and styrene into the STM
chamber through a variable leak valve. Dissolved
atmospheric gases in the styrene and 4-methyl-styrene were
expelled from the liquid phase samples by repeated
freeze-pump-thaw cycles. Prior to dosing, the STM tip was
retracted $\sim$ 1 micrometer from the H:silicon surface.
During dosing, the flow of vapour phase reactant was
adjusted to bring the chamber pressure to $\sim$
4$\times$10$^{-7}$ Torr, and held for several tens of
seconds (integrated doses given below).  After dosing
(valve shut), the chamber pressure returned below
1$\times$10$^{-10}$ Torr.

Images of the H:Si(100) surfaces were collected in UHV at
room temperature using an Omicron STM1. Electrochemically
etched tungsten tips were cleaned in-situ by electron beam
bombardment and field emission prior to STM imaging.
Constant current topographic STM images were acquired
using a constant tunnel current of 60 pA. Quoted
dimensions along the lateral and vertical directions have
been scaled to agree with known values for the silicon
dimer spacing, and silicon terrace height, respectively.
4-methyl-styrene/styrene heterostructures were grown on
several H-silicon surfaces, and studied with different
STM tips.

Figs. \ref{Fig_1} and \ref{Fig_2} show the growth and
bias-dependent filled-state imaging of a
methyl-styrene/styrene heterostructure.
Fig.\ref{Fig_1}(a), acquired using a sample voltage
(V$_s$) of -3.0 V, shows the formation of the first line
segment following a 8 Langmuir dose of methyl-styrene (1
Langmuir = $10^{-6}$ Torr$\cdot$seconds). Approximately 8
molecules of methyl-styrene have reacted at the indicated
location. The bright feature at the end of the line
segment (indicated by the white arrow) corresponds to the
terminal silicon radical (i.e. a bare silicon atom which
has transferred its hydrogen atom to the adjacent
methyl-styrene molecule). Its presence indicates that the
reaction between the methyl-styrene and the silicon
surface at this particular site has not reached saturation.

Fig.\ref{Fig_1}(b) shows the same region of the crystal
following a 78 Langmuir exposure of styrene. Starting at
the silicon radical in (a), approximately 8 molecules of
styrene have subsequently reacted to form a
methyl-styrene/styrene heterostructure. The location of
the heterointerface is indicated by the white wedge. The
silicon radical present in (a) is absent from the end of
the styrene line segment in panel (b), and in panels (c)
and (d) acquired at lower magnitude bias. This suggests
that linegrowth has terminated at an unresolved surface
defect (such as a dihydride site), or has been capped by a
H-atom deposited by the STM tip.\cite{Other}

Panels (b) - (d) show the topographic envelope of the
heterostructure at varying sample bias. At $V_s$ = -3.0 V
[panel (b)], the constant current contours follow a
relatively smooth envelope modulated by the physical
height of the methyl-styrene and styrene line segments
above the H:silicon surface. At this value of the bias 
the aromatic $\pi$ states in the styrene and
methyl-styrene lie above the tip Fermi-level.  The tunnel
current therefore results from electrons which tunnel
from underlying occupied states within both the silicon
and surface bound molecules into empty states within the
STM tip.  The extent to which molecular orbital states
emptied by tunneling are replenished directly by
underlying silicon states (i.e. vertical transport across
molecules) or indirectly by nearest neighbour molecules
(reflecting a degree of lateral transport across the
molecular chain) is considered in the following theory
sections.

In panel (c) [V$_s$ = -2.4 V], the aromatic $\pi$ states
move below the tip Fermi-level, and the imaging current is
supplied by silicon states which tunnel across the surface
bound molecules. In this regime, the methyl-styrene and
styrene image with similar height. Also, molecular scale
corrugation appears within the topographic envelope. In
panel (d) [V$_s$ = -1.8 V], the molecular corrugation
shows different contrast between the methyl-styrene and
styrene line segments.
  
To better resolve differences in the imaging
characteristics between both segments, 0.2 nm wide
topographic cross-sections were extracted along two
parallel axes within the heterostructure envelope. Images
were registered using a least squares fitting algorithm to
allow extraction of cross-sections from nominally
identical locations between images acquired at different
bias.

Cross-sections taken along the direction of the blue arrow
(Fig.\ref{Fig_1} panel (d) inset) appear in panel (a) of
Fig.\ref{Fig_2} and are centred above the chemical
attachment points between the heterostructure and the
underlying dimer row. Cross-sections taken along the
direction of the red arrow, appear in panel (b) of
Fig.\ref{Fig_2}, and are centred to the right of the
trench separating the underlying dimer rows. 
Cross-sectional data extracted from images (not shown)
acquired at sample voltages of -2.0 V, -2.2 V, -2.6 V, and
-2.8 V have been added to the data presented in
Fig.\ref{Fig_2}.

In panels (a) and (b) of Fig.\ref{Fig_2}, the height
envelope corresponding to the methyl-styrene line segment
is situated between 2.5 nm and 6.0 nm along the distance
axis.  The styrene line segment is situated between 6.0 nm
and 9.8 nm.  The peaks centred at 1.5 nm (in both panels)
and at 10.3 nm (in panel (a) only) correspond to the
isolated (and unidentified) structures present in the
upper and lower left hand corners of the images in
Fig.\ref{Fig_1}.  

In each panel the position of the corrugation maximum
associated with the interfacial methyl-styrene molecule is
indicated by a black arrow.  The position of the chemical
interface in the heterowire is made evident by the
differential height contrast observed between the two line
segments at high and low bias. Superposition of the
heterowire envelope with registered cross-sections
extracted from Fig.\ref{Fig_1} (a) (not shown) agree with
this determination.  In this case, the topographic maximum
(i.e. the silicon radical) at the end of the
methyl-styrene segment in Fig.\ref{Fig_1} (a) lies
between the as-identified interfacial methyl-styrene and
styrene corrugation maxima in the completely grown
heterowire.  Other methyl-styrene/styrene heterowires
studied displayed similar agreement between the apparent
and expected locations of the heterojunction.

The main features displayed in Fig.\ref{Fig_1} are visible
in the cross-sectional data presented in panels (a) and
(b) of Fig.\ref{Fig_2}.  As the magnitude of the imaging
bias is increased, the apparent height of the
methyl-styrene and styrene line segments increases and
begins to saturate (with methyl-styrene imaging above
styrene).  At low magnitude bias (where the contribution
of molecular $\pi$ states to the tunnel current is
suppressed), the molecular corrugation is more apparent.  

While in panel (b) the height of the methyl-styrene line
segment is comparable to (at low magnitude bias) or
greater than (at high bias) that of styrene, a different
behaviour is observed along the chemical attachment points
in (a). At low magnitude bias, $|V| < 2.6$ V, the
methyl-styrene molecules image lower than the styrene
molecules.  Given that methyl-styrene extends farther
beyond the H:silicon surface than styrene, and the
similarity in molecular conformation anticipated for the
molecules in each segment, this observation seems
unexpected.

At the heterojunction, the imaging contrast between the
interfacial molecules (styrene and methyl-styrene) appears
to a first approximation comparable to the contrast
resolved for these same molecules within their respective
line segments.  A significant departure from this
behaviour is observed at elevated magnitude bias 
($|V|{\raisebox{-0.6ex}{$\,\stackrel
{\raisebox{-.2ex}{$\textstyle >$}}{\sim}\,$}}2.6$ V),
where the interfacial methyl-styrene appears lower than
the molecules located within the line segment.  The
molecules at either end of heterowire also image with
decreased height.  Finally, a slight upward bowing
(roughly centred at the heterointerface) appears
superimposed on the bulk of the cross-sectional curves.

In the absence of electronic interaction between molecules
and geometrical differences between them, every molecule
along a homomolecular wire (or a homomolecular line
segment in the case of a heterowire) would image with
identical properties.  This is not observed.  The
implication is that the observed height profiles along the
heterowire reflect some combination of: i) conformational
changes in the molecules along the chain due to
intermolecular forces, ii) dispersion interactions
leading to differential broadening and shifting of
molecular electronic levels along the chain, and/or iii)
tunneling current collected by the STM tip above a given
molecule receiving contributions from neighbouring
molecules along the chain reflecting some degree of
lateral current transport across the stacked aromatic
rings.  

To determine the origin of the observed contrast in
molecular corrugation, and the topographic height
variations resolved along the heterowire structure, a
theoretical model for the heterowire is developed and its
current transport properties are considered in sections IV
to VIII.  

\section{Modeling the Structure}
\label{structure}

The spacing between the tops of the molecules seen in our
experimental STM images of styrene/methyl-styrene chains
on the (001) silicon surface varies along the length of
the chain and is larger near the ends of the chain than
near its center. Because of the variable mismatch between
the molecular spacing and the underlying silicon lattice,
it is not appropriate to model these molecular chains
theoretically as structures periodic along the length of
the chain with a small unit cell. They are therefore
modelled here as finite clusters such as that depicted in
Fig.\ref{Fig_3} which shows a molecular chain with a
heterojunction, consisting of four styrene and four
methyl-styrene molecules. The molecules bond to a
dimerized Si (001) surface that is represented by two
rows each of 10 silicon dimers together with four
underlying atomic layers of silicon atoms. All of the
dangling bonds are passivated with hydrogen. The lowest
carbon atom of each molecule bonds to a surface Si atom
and the molecules lean over the ``trench'' between the
two rows of silicon dimers.\cite{reversebonding}

A complete {\it ab initio} relaxation of this cluster 
using Kohn-Sham density functional theory would be
impractical, so the following  hybrid approach was
adopted to calculate its approximate structure: An {\it
ab initio} density functional relaxation was carried
out\cite{Gaussian} of a smaller cluster consisting of one
styrene and one methyl-styrene molecule over two rows
each of two silicon dimers, with two additional
underlying layers of silicon, all of the dangling bonds
passivated with hydrogen. This small cluster shown in
Fig.\ref{Fig_4} is similar to the immediate vicinity of
the junction between styrene and methyl-styrene chains in
Fig.\ref{Fig_3}. The relaxation of this small cluster was
carried out keeping the atoms of the two lower silicon
layers fixed at positions corresponding to a bulk silicon
crystal lattice but allowing the positions of all of the
atoms of the molecules, the silicon dimer atoms and all of
the hydrogen atoms to relax freely. The main difference
between the relaxed geometries of the two molecules in
Fig.\ref{Fig_4} and that of an {\em isolated} molecule
bound to the same silicon site is that the two molecules
tilt somewhat away from each other. This tilt occurs
mainly through changes in the sequence of dihedral angles
defined (in the Z-matrix description of the system) by the
orientations of the bonds connecting Si atom 1 to C atom
2, C atom 2 to C atom 3, and so on along the carbon atom
chain through C atom 6 of the styrene molecule (see
Fig.\ref{Fig_4}), and of the corresponding bonds of the
methyl-styrene molecule. Thus the relaxed geometry of the
cluster in Fig.\ref{Fig_3} was generated by assigning
initially to all of the atoms of the molecules positions
derived from the relaxed geometries of the respective
atoms of the smaller cluster in Fig.\ref{Fig_4}. Then
this initial structure was relaxed so as to minimize the
energy of the cluster using the AM1 empirical
model,\cite{AM1} but allowing {\em only} the above
dihedral angles, the dihedral angles defining the
orientations of the C-H bonds involving carbon atoms 2 and
3 and corresponding C-H bonds on other molecules, and the
dihedral angles defining the orientation of the triplets
of H atoms belonging to the methyl groups to vary. During
the AM1 relaxation a further constraint was also imposed,
namely, that the tilts of the styrene and methyl-styrene
molecules at the ends of the 8 molecule chain be such
that the average lateral spacing between the top carbon
atoms of the molecules approximate the experimentally
observed average spacing of the molecules near the
junction of the styrene and methyl-styrene
chains.\cite{just} The final model geometry of the 8
molecule chain is thus representative of the region of a
much longer molecular chain around the heterojunction. A
similar procedure was used to obtain the relaxed geometry
of the chain of 8 styrene molecules used in the
theoretical studies of lateral electron transport along
the molecular chain that are discussed in  Section
\ref{lateral}. However, the structures used in our
theoretical studies of the ends of the molecular chains
were obtained somewhat differently as is described in
Section
\ref{ends}. Finally, it was found that the energy of the
structure with methyl-styrene molecules varies by much
less than $kT$ (where $T=$ room temperature) when the
methyl groups are rotated about the bonds connecting them
to the benzene rings of the molecules. Since the
experiments on these systems were carried out at room
temperature, the calculated STM tip currents presented
below are averages over orientations of the methyl
groups. 

The atomic structure of the tungsten STM tip in our
experiments is unknown and the tip was modelled
arbitrarily as a (001)-oriented (relative to the (001) Si
surface) clean bcc tungsten tip terminating in a single
tungsten atom. A total of 15 tungsten atoms arranged in
layers of 1,4,5,4 and 1 atoms along the (001) direction
were included in the model of the tip. These atoms were
placed at their nominal positions in a bulk bcc tungsten
lattice except for the terminal atom whose position was
relaxed along the (001) direction (towards the center of
the tungsten cluster) so as to minimize the Kohn-Sham
energy of the 15 atom tungsten
cluster.\cite{Gausstip}          The effect of an
alternate tip on the calculated STM currents, obtained by
replacing the apex atom with a single silicon atom whose
position was relaxed appropriately, was also investigated.
   
\section{Modeling the Electronic Structure}
\label{electronicstruct}
 
Most theoretical work on electronic transport in
molecular  wires with metal contacts has been based on 
semi-empirical tight-binding models or {\it ab initio}
Kohn-Sham density functional calculations of the
electronic
structure.\cite{Datta,metalcontacttheory1,Emberly,metalcontacttheory2,Kushmeric,metalcontacttheory3,Guo}
However, the atomic structures of the molecule-metal
junctions are not known. Thus, with a few
exceptions,\cite{Guo,Lee} comparison of theoretical
predictions with experiments on metal/molecule/metal
wires is subject to large uncertainties. By contrast, the
bonding geometries of many small organic molecules to
silicon substrates are known,\cite{Wolkow}  as are the
specific atomic silicon sites involved in the
carbon-silicon bonds anchoring styrene and methyl-styrene
molecular chains to the
silicon.\cite{Lopinski,Cho,Hofer,Wang} Thus comparison
between theory and experiment is subject to much less
uncertainty for these molecules on silicon than for
metal/molecule/metal wires. This allows us to assess the
validity of different models of the electronic structure
of the molecular chains on silicon by comparison with
experimental data.  

Previous theoretical work modeling the current-voltage
characteristics of single styrene molecules on silicon
found {\it ab initio} Kohn-Sham density functional
calculations unsuitable for treating the electronic
structure of the silicon substrate.\cite{Rakshit} The
present work on styrene and methyl-styrene chains on
silicon reached a similar conclusion: Our density
functional calculations that employed localized orbital
bases appropriate for quantum chemistry\cite{Gaussian}, in
common with previous density functional calculations for
styrene chains on silicon that employed plane wave
bases,\cite{Hofer} yielded molecular HOMO levels aligned
with the highest occupied silicon valence states. This
level alignment is unrealistic; experimentally the
molecular HOMO energy is considerably lower. Furthermore, 
if it were correct then STM images of the styrene molecules
probing occupied states at low bias would show a current
node at the center of each molecule since the molecular
HOMO is a
$\pi$-like orbital on the benzene ring. However, this is
not seen experimentally. Finally, density functional
calculations in the local density approximation seriously
underestimate the band gap of {\em bulk} silicon.   For
these reasons\cite{DFTtouble} a different approach based on
extended H{\"u}ckel theory will be adopted here in modeling
the electronic structure of the molecules, the silicon
substrate and the tungsten STM tip. Our model corrects the
above deficiencies of density functional theory and yields
current {\em maxima} over the centers of the molecules for
low negative applied substrate bias, consistent with the
nature of the corrugation along the molecular chain that is
observed experimentally.

Extended H{\"u}ckel theory is a semi-empirical tight
binding scheme from quantum chemistry that provides an
approximate description of the electronic structure of
many molecules. It describes molecular systems in terms
of a small set of    Slater-type atomic orbitals $\{
|\phi_i\rangle \}$, their overlaps $S_{ij} =
\langle\phi_i | \phi_j\rangle$ and a Hamiltonian matrix
$H_{ij} =
\langle\phi_i |H| \phi_j\rangle$. The diagonal
Hamiltonian elements $H_{ii} = \epsilon_i$  are chosen to
be the atomic orbital ionization energies. In the
Wolfsberg-Helmholz form of the model, the non-diagonal
elements are approximated by 
\begin{equation}
H_{ij} = K
S_{ij}(\epsilon_i + \epsilon_j)/2
\label{Hij}
\end{equation}
where $K$ is a phenomenological parameter usually chosen
to be 1.75 for consistency with experimental molecular
electronic structure data. The extended H{\"u}ckel model
has been used successfully to explain the experimental
current-voltage characteristics of a variety molecular
wires connecting metal
electrodes.\cite{Datta,Emberly,Kushmeric} However,
without modification it is unsatisfactory for crystalline
silicon as it predicts a band gap that is direct and
roughly three times larger than that found experimentally
for this material. In the present work these deficiencies
of the extended H{\"u}ckel model have been corrected by
replacing equation (\ref{Hij}) for the Hamiltonian matrix
elements between the atomic orbitals of silicon by 
\begin{equation}
H_{ij} = K_{ij}
S_{ij}(\epsilon_i + \epsilon_j)/2
\label{Kij}
\end{equation}
where the $K_{ij}$ are fitting parameters\cite{fit}
chosen so that the modified extended H{\"u}ckel model
obtained in this way yields the correct silicon band
structure. The extended H{\"u}ckel Hamiltonian matrix
elements for tungsten were adjusted similarly so as to
match the band structure of bcc tungsten,\cite{pap} 
however in the case of tungsten some modification of the
extended H{\"u}ckel model's orbital energies
$\epsilon_i$ was also found to be necessary to achieve a
good fit. Finally a realistic alignment between the
molecular HOMO level and the Fermi levels of the silicon
and tungsten clusters was achieved by shifting the silicon
and tungsten orbital energies $\epsilon_i$ by appropriate
amounts
$E_{Si}$ and $E_{W}$ respectively.\cite{pap} Because of
the non-orthogonality of the extended H{\"u}ckel basis
states, such shifts of the diagonal matrix elements of the
Hamiltonian also require\cite{shift} corresponding shifts
$\Delta H_{ij}$ of the non-diagonal elements
$H_{ij}$, which were taken to be  
\begin{equation}
\Delta H_{ij} =
S_{ij}(E_i + E_j)/2
\label{gauge}
\end{equation}
 where $E_k = E_{Si}$ or $E_{W}$ if $\{ |\phi_k\rangle
\}$ is a silicon or tungsten orbital, respectively.

Fig.\ref{Fig_5} shows the key features of the electronic
structure of the eight-molecule styrene/methyl-styrene
heterojunction on the hydrogen passivated silicon
cluster  whose geometry is depicted in Fig.\ref{Fig_3},
as described by the above electronic structure model: The
energy eigenvalues and eigenstates of the model
Hamiltonian were calculated and a Mulliken analysis was
carried out to determine the silicon, carbon and hydrogen
content of each eigenstate. The carbon content was
further resolved according to whether the carbon atom
belongs to a styrene or methyl-stryrene molecule. The
results are shown as histograms in Fig.\ref{Fig_5} a
(Si), b (C/styrene), c (C/methyl-styrene) and d (H),
summed for the sake of clarity over the eigenstates in
bins of energy of width 0.1 eV. The vertical line labelled
``Si HOMO'' marks the energy of the highest occupied state
of the cluster which has mainly silicon content. The
carbon band below about -12.3 eV corresponds to molecular 
HOMO orbitals while the carbon band around -8.2 eV
corresponds to the molecular LUMO. Because the HOMO
(LUMO) of ethylbenzene is close in energy and generally
similar to  the HOMO (LUMO) of methylethylbenzene, the
molecular HOMO (LUMO) bands due to the styrene and
methyl-styrene on silicon (Fig.\ref{Fig_5}b and c) are
similar to each other although the styrene band edges are
slightly lower (by $\sim 0.1$-0.2 eV) in energy than those
for methyl-styrene.

Note also the weak carbon features between the molecular
HOMO band edge at -12.3 eV and the (mainly silicon)
cluster HOMO at -11.3 eV. These are almost identical for
the styrene and methyl-styrene molecules and  are due to
electron tunneling through the molecules from occupied
states of the silicon substrate that lie above the
molecular HOMO band. This tunneling is responsible for
the STM current for low negative sample bias at which the
experimentally observed STM images of the molecules are
similar to those calculated using the present model
(Fig.\ref{Fig_6} plot L in Section \ref{vertical}) but
differ markedly from the those predicted by density
functional calculations as was discussed
above.              

Since the styrene and methyl-styrene molecules are
coupled more strongly to the silicon substrate than to
the STM tip, it is assumed for simplicity in the present
work that when a bias voltage $V$ is applied between the
STM tip and sample, the entire voltage drop occurs between
the molecules and the STM tip.\cite{drop} The
corresponding shift $eV$ of the energy levels of the tip
relative to the sample is then included in $E_{W}$, and
thus also contributes to the off-diagonal matrix elements
of the Hamiltonian through equation (\ref{gauge}). 

Finally, it should be emphasized that because the present
model includes only a few layers of silicon atoms, the
electronic structure of the silicon that it describes is
characteristic of the immediate vicinity of the silicon
surface only and does not include band bending effects
that occur over larger length scales within the silicon.
Furthermore the bias voltages considered in the model are
those between the region slightly below the silicon
surface and the STM tip, whereas bias voltages measured
experimentally are those between the STM tip and the deep
silicon interior that is separated from the surface by
regions where significant bias-dependent band bending
occurs. Thus from the stand point of theory the applied
bias will be characterized according to the location of
the Fermi level of the tungsten tip (or, more precisely,
its electrochemical potential) relative to the features
of the carbon and silicon partial densities of states
that are shown in Fig.\ref{Fig_5}, rather than the
numerical value of experimental bias voltage.         

\section{Modeling the electronic transport}
\label{transport}

In the transport calculations reported here the
electronic structures of the silicon substrate, molecules
and tungsten tip are all treated together as a single
system described by a single Hamiltonian. Thus, unlike in
STM image calculations based on Bardeen tunneling theory,
here the Hamiltonian matrix elements connecting orbitals
of the molecules and those of the tip are {\em not}
assumed to be small. Because of this the STM current can
be calculated for small separations between the tip and
molecules as well as large. 

The calculations of the current are based on Landauer
theory\cite{book} which relates the electric current $I$ 
through a nanostructure under an applied bias voltage $V$
to the multichannel  electron transmission probability
$T(E,V)$ through the nanostructure at energy $E$ as\cite{TEV}
\begin{equation}
I(V) = \frac{2e}{h} \int_{-\infty}^{\infty}
dE\:T(E,V)\left( f(E,\mu_{s}) - f(E,\mu_{d})\right)
\label{Landauer}
\end{equation}
where $f(E,\mu_{i}) ={1}/{(\exp[(E-\mu_{i})/kT] + 1)}$
and $\mu_{i}$ is the electrochemical potential of the
source ($i=s$) or drain ($i=d$) electrode.\cite{temp}

In the present work the source and drain electrodes are
modelled as arrays of ideal leads:  For calculations of
vertical transport in the STM geometry, the electrode
connected to the silicon substrate is modeled as an array
of 152 ideal single-channel leads, one such lead coupled
to the 1$s$ orbital of each of the hydrogen atoms that
passivate the silicon dangling bonds of the three lower
layers of silicon atoms of the silicon cluster that
represents the silicon substrate in Fig.\ref{Fig_3}. The
electrode connected to the STM tip was modelled as an
array of 81 ideal single-channel tight-binding leads, one
such lead coupled to each of the 9 ($s,p,d$) valence
orbitals of each atom of the model tungsten tip that is
not the tip atom itself or one of its 5 closest neighbors.
As well as mimicking macroscopic electrodes by supplying
an ample electron flux to the system, this large number
of ideal source and drain leads has a similar effect to
phase-randomizing B{\"u}ttiker
probes\cite{Buttikerprobes} in minimizing the influence
of dimensional resonances due to the finite sizes of the
tungsten and silicon clusters employed in the model. Each
ideal lead $i$ was modelled as a semi-infinite
tight-binding chain with a single orbital per site, a site
energy
$\alpha_i$ and nearest neighbor hopping matrix element
$\beta$. $\alpha_i$ was chosen to be equal to the energy
(including the shift, if any, due applied bias described
in Section
\ref{electronicstruct}) of the hydrogen or tungsten
orbital to which lead $i$ was coupled. The hopping matrix
elements $\beta$ were all taken to have a magnitude of
5 eV, sufficiently large that the eigenmodes of all of the
ideal leads in the energy window between $\mu_{s}$ and
$\mu_{d}$ in eq. (\ref{Landauer}) be
propagating.\cite{temp} The coupling matrix elements
$W_i$ between the ideal leads and their respective
hydrogen and tungsten orbitals were also set equal to
$\beta$.

To calculate $T(E,V)$ and hence evaluate
eq.(\ref{Landauer}), the transformation to a different
Hilbert space described in Refs. \onlinecite{orthog1} and
\onlinecite{orthog2} was made, mapping the non-orthogonal
basis of atomic orbitals discussed in Section
\ref{electronicstruct} to an orthogonal basis. The
Lippmann-Schwinger equation 
\begin{equation}
|\Psi_{i}\rangle = |\Phi_{o,i}\rangle + G_{o}(E) W
|\Psi_{i}\rangle
\label{eq:LS}
\end{equation}
describing electron scattering between the source and
drain leads via the tungsten tip, molecules and hydrogen
terminated silicon cluster was solved numerically for
$|\Psi_{i}\rangle$ in the alternate Hilbert
space.\cite{orthog1,orthog2} In eq.(\ref{eq:LS}) 
$G_{o}(E)$ is the Green's function for the decoupled
system (i.e., with the coupling  between the ideal leads
and the hydrogen and tungsten orbitals switched off),
$|\Phi_{o,i}\rangle$ is the eigenstate of the decoupled
ideal source lead $i$ with energy
$E$ and
$|\Psi_{i}\rangle$ is the corresponding scattering
eigenstate of the complete system with the coupling $W$
between the ideal leads and the hydrogen and tungsten
orbitals switched on. The scattering amplitudes $t_{ji}$
from the ideal source lead $i$ to drain lead $j$ at energy
$E$ were extracted from the scattering eigenstates
$|\Psi_{i}\rangle$ and the transmission probability that
enters eq. (\ref{Landauer}) was then calculated from
\begin{equation}
T(E,V) = \sum_i \sum_j \left | \frac{v_j}{v_i} \right |
|t_{ji}|^2
\label{eq:multiT}
\end{equation}
where ${v_i}$ and ${v_j}$ are the electron velocities in
ideal leads $i$ and $j$ respectively at energy $E$.

Lateral transport through chains of styrene molecules on
the silicon substrate was modelled similarly. However, in
this case no STM tip was included in the model and no
ideal leads were coupled to the silicon cluster or to the
hydrogen atoms passivating it. Instead, 18 ideal
single-channel leads representing a source electrode were
coupled directly to the styrene molecule at one end of the
styrene chain, one ideal lead coupling to each of the
atomic valence p-orbitals of each carbon atom of the
benzene ring of the molecule. 18 ideal single-channel
leads representing the drain electrode were coupled
similarly to the styrene molecule at the other end of the
molecular chain. In this case transport through the
styrene chains via bands derived from molecular HOMO and
LUMO orbitals was being investigated so the site energies
$\alpha_i$ of the orbitals making up the ideal leads were
chosen lying within the molecular HOMO and LUMO energy
bands, respectively. The nearest neighbor hopping matrix
element
$\beta$ of the ideal leads was chosen so that the widths of
the energy bands of the ideal leads were comparable to the
widths of the energy bands derived from the molecular
HOMO and LUMO energy levels.

\section{Results: Vertical transport in the STM Geometry}
\label{vertical}

\subsection{Negative sample bias}
\label{negative}
Fig. \ref{Fig_6} shows representative results for the
calculated current flowing between the tungsten STM tip
and the styrene/methyl-styrene molecular chain on silicon
that is depicted in Fig.\ref{Fig_3}.  The silicon
substrate is biased negatively with respect to the STM
tip so that occupied states of the molecular chain and
substrate are being probed. Each curve corresponds to a
scan of the STM tip along the line of molecules at a
constant height above the silicon substrate.\cite{traj}
The lateral position of the highest carbon atom of  each
molecule is indicated by a dotted vertical line labelled m
(methyl-styrene) or s (styrene). Curves L and L$'$
correspond to a low value of the bias such that the Fermi
level of the tungsten tip is at -12.0 eV as indicated by
the arrows labelled L in  Fig.\ref{Fig_5} (b) and (c).
Since in this case the tungsten Fermi level is well above
the upper edge of the molecular HOMO band in
Fig.\ref{Fig_5} the current is due to electron {\em
tunneling} through the molecules from the silicon
substrate states to the STM tip.  Curve H is for a higher
bias with the tip Fermi level at -12.6 eV as indicated by
the arrow labelled H in  Fig.\ref{Fig_5}(b). In this case
the tungsten Fermi level is within of the molecular HOMO
band that is located on the molecular benzene rings and
mediates transport between the silicon substrate and the
tip. For curves L and H the tungsten atom at the end of
the tip is at a height 2 {\AA} higher than the highest
carbon atom of the molecular chain while for curve L$'$
the tip atom is 3 {\AA} higher than the highest carbon
atom. (Note that for curve H the current shown has been
scaled down by a factor of 10 for clarity).

Curve L is similar to experimental scans along the
molecular chain at low negative sample bias
(Fig.\ref{Fig_2} (a) and (b)),  although the
experimental scans were taken at constant current
while the calculation was performed at constant tip
height:\cite{heightvscurrent} In curve L each
molecule appears as a distinct current {\em maximum}
consistent with the experimental data,\cite{nonode} 
and the local current over each styrene molecule is
somewhat stronger than over a methyl-styrene
molecule  (as in Fig.\ref{Fig_2} (a)) despite the
fact that the methyl-styrene molecule is the taller
of the two.\cite{strongcoupl} Interestingly, as the
tip is raised higher above the molecular chain at
constant (low) bias the current flowing through the
styrene molecules is predicted to decrease initially
much faster than that through the methyl-styrene,
resulting in the significantly weaker styrene current
seen for curve L$'$. We attribute this to different
behavior of weak transport resonances (involving both
tip states and carbon-silicon interface states) for
the styrene molecule than for the methyl-styrene
molecule that is much closer to the tip and thus
interacts more strongly with it.\cite{othertips}

As the bias increases, the calculated current through the
methyl-styrene molecules increases more rapidly than
through the styrene and the relative amplitude of the
current corrugation along the molecular chain decreases
along both the styrene and methyl-styrene parts of the
chain. Both of these effects (that are clearly visible in
curve H) are also seen experimentally in Fig.\ref{Fig_2}
(a) and (b).\cite{othertips} 

The above theoretical results imply  that the
observation of an experimental STM profile similar to L
in Fig. \ref{Fig_6}  signals that the STM tip Fermi
level lies between the Si valence band edge at the Si
surface and the HOMO band of the molecular chain,
whereas the observation of a profile similar to H
signals that the tip Fermi level lies below the upper
edge of the HOMO band. Since a transition from an L
type profile to a H type profile is in fact observed
experimentally as the magnitude of the substrate bias
is increased (see Fig. \ref{Fig_2}), our experimental
and theoretical findings taken together confirm that
the molecular HOMO band does in fact lie significantly
below the silicon valence band edge, as we have already
suggested in Section
\ref{electronicstruct} based on the nature of the 
corrugation observed along the molecular chain at low
negative substrate bias.

Another interesting aspect of the results shown in Fig.
\ref{Fig_6} occurs at the junction of the methyl-styrene
and styrene chains: At low bias (curves L and L$'$) the
current profile of the styrene molecule at the junction
is quite similar to those of the styrene molecules far
from the junction, and the same is true of the
methyl-styrene molecules. However, this is {\em not} true
at high bias (curve H)  where the current decreases {\em
monotonically} from the methyl-styrene side of the
junction all the way through the location of the first
molecule on the styrene side; the first styrene molecule
is not visible as a distinct entity, in marked contrast to
what is seen near the junction in curve L$'$. We
attribute this smearing of the boundary between the
methyl-styrene and styrene chains in the STM image at
high bias to delocalization of electron states along the
molecular chain associated with overlapping molecular
HOMO levels on adjacent molecules. This effect is not
found in the STM images at low bias (curves L and L$'$)
where the states outside of the molecular HOMO band are
responsible for conduction.  We will return to this topic
in our discussion of positive sample bias below  and its
relation to lateral transport along the molecular chain
will be considered at the end of Section \ref{lateral}.

Increased smearing of the boundary between the
methyl-styrene and styrene chains with increasing bias is
also seen in the experimental STM data in Fig.\ref{Fig_2}
(a) and (b). However, experimentally the apparent height
of the molecular chain varies on both sides of the
junction but more so on the methyl-styrene side. The
reason for this difference between theory and experiment
is unclear at the present time. 

The calculated current peaks for STM tip scans  at low
bias and constant height in the direction orthogonal to
that of the scans in Fig. \ref{Fig_6} exhibit internal
structure. This can be seen in Fig.\ref{Fig_7}(a) where
curve sL (mL) is the result of such a scan across the
third styrene (methyl-styrene) molecule from the junction
for the same bias and tip height as scan L in
Fig.\ref{Fig_6}. With increasing bias voltage the
internal structure becomes less prominent and the scans
evolve to a simpler single-peak structure.

\subsection{Positive sample bias} 
\label{positive}
           
Fig. \ref{Fig_8} shows the calculated current between the
STM tip and the  molecular chain for some values of
positive substrate bias. The tip is scanned along the
molecular chain at constant height,\cite{traj} 2 {\AA}
higher than the highest carbon atom in each case. The
silicon substrate is biased positively with respect to
the STM tip so that unoccupied states of the molecular
chain and substrate are being probed. Dotted vertical
lines again indicate positions of the highest carbon
atoms of the styrene (s) and methyl-styrene molecules.
The tungsten Fermi level is located below the molecular
LUMO band for curves ll and l, slightly above the lower
edge of the LUMO band for curve e, and well inside the
LUMO band for curve h, at -9.3 eV, -8.6 eV, -8.39 eV and
-8.2 eV respectively, as is indicated by arrows in Fig.
\ref{Fig_5}(b). At low positive bias (curves ll and l)
electrons tunnel from the tungsten tip to the silicon
substrate states through the molecules {\em outside} of
the energy range of the molecular LUMO band and the
current shows a peak over each molecule, as for the case
of low negative substrate bias (curves L and L$'$ of Fig.
\ref{Fig_6}). As the bias increases and the tungsten
Fermi level begins to enter the molecular LUMO band, the
current develops a minimum over the center of each
styrene molecule because the molecular LUMO has
$\pi$-like character with a wave function node in the
plane of the benzene ring.\cite{moreaboutdips} A similar
effect also happens over the methyl-styrene molecules
with increasing bias, but more slowly because of the
presence of the methyl group between the benzene ring
and the STM tip: The amplitude of the current
corrugation over the methyl-styrene chain decreases from
curve ll to curve e and then reverses so that for curve
h the current is depressed over the center of each
molecule. The amplitude of the current corrugation also
decreases with increasing bias (from curve e to curve h)
over the styrene chain. At low bias the contrast between
the methyl-styrene and styrene decreases with decreasing
bias (from curve l to curve ll) but does not reverse as
in Fig.\ref{Fig_6}. 

Interestingly, the current for curve e develops a strong
overall slope from molecule to molecule extending across
the {\em entire} styrene chain, an effect that is not
found for curves ll, l and h or under negative substrate
bias (Fig. \ref{Fig_6}). There is also a weaker slope
across the  methyl-styrene chain. This suggests that
electronic communication between the molecules  over a
length of several inter-molecular spacings is important
at this bias voltage and strongly affects the calculated
STM image of the molecular chain. This will be explored
further in   Section \ref{lateral}. 

Fig.\ref{Fig_7}(b) shows the calculated tip current for
scans across the molecular chain at positive substrate
bias in the direction orthogonal to the scans in Fig.
\ref{Fig_8}. Curve se (me) is the result of such a scan
across the third styrene (methyl-styrene) molecule from
the junction for the same bias and tip height as scan e in
Fig.\ref{Fig_8}. Note the double peaked structure in scan
se. This double peaked structure over styrene evolves
into a single peak with increasing and decreasing bias,
while the scan across methyl-styrene remains
single-peaked.

\section{Lateral Transport and its Influence on STM Images}
\label{lateral} 

The overlaps between the molecular HOMO and LUMO states
of adjacent styrene and methyl-styrene molecules in the
molecular chains on silicon  have the potential to give
rise to bands of electronic states delocalized along the
molecular chains.\cite{Cho,Hofer} However, the  molecular
LUMO (HOMO) energy levels are degenerate with states of
the silicon  conduction (valence) band and clearly must
hybridize with them to some degree. Moreover these
molecular energy levels correspond to excited states of
the total system and electrons (holes) occupying them
must eventually decay to lower energy conduction
(valence)  states of the substrate due to inelastic
processes. However, it is still interesting from a
theoretical perspective to determine how much the
hybridization between a molecular chain's LUMO and HOMO
bands and the silicon conduction and valence bands
affects electron or hole transport along short stretches
of the molecular chains. For example, if an electron
(hole) is injected into the LUMO (HOMO) orbital of a
molecule belonging to the chain and extracted by a probe
at another site along the molecular chain, is the
transport between the two sites dominated by conduction
through the molecular chain or through the substrate or
do these two conduction channels make comparable
contributions to the transport? A related quantity, the
characteristic distance that an electron or hole injected
into a LUMO or HOMO state of the molecular chain from an
STM tip travels along the chain before being absorbed into
the silicon substrate, may influence the STM image of a
molecular chain, particularly near a heterojunction. In
this Section the issue of parallel transport along the
molecular chain and through silicon substrate over
nanoscale distances is addressed theoretically, and the
results are related qualitatively to the findings
regarding the calculated STM images of the molecular
chains that were presented in Section
\ref{vertical}. 

The approach used is to calculate the multichannel
Landauer transmission probabilities for electrons between
a pair of ideal electrodes coupled to the ends of a
molecular chain of 8 styrene molecules on a dimerized
hydrogen-terminated silicon substrate. For the
calculations presented here 9 layers of silicon atoms
were included in the cluster modeling the silicon 
substrate. However, results obtained with a 5 silicon
layer model of the cluster were similar. The ideal source
and drain electrodes were coupled to the benzene rings of
the molecules at the two ends of the styrene chain and
were represented by arrays of ideal single channel leads
as described at the end of Section \ref{transport}. The
energy bands of the ideal leads were taken to be centered
at $\alpha_i = -12.7$ eV (-8.25 eV) for studies of the
lateral transport in the energy range of the molecular
HOMO (LUMO) band; the ideal lead hopping parameter
$\beta$ was chosen to be -0.25 eV for an ideal lead band
width of 1 eV. The hamiltonian matrix elements coupling the
ideal leads to the carbon p-orbitals of the molecular
benzene rings was chosen to be $0.4 \beta$.

The calculated Landauer transmission probability $T(E)$
(per spin channel) through the system for energies
corresponding to the LUMO band of the styrene chain is
shown in Fig.
\ref{Fig_9}(a). These results show that the system is a
fairly good one-dimensional conductor with at least two
partially-transmitting channels per spin in the energy
range around the middle of the LUMO band since 
$1{\raisebox{-0.6ex}{$\,\stackrel
{\raisebox{-.2ex}{$\textstyle <$}}{\sim}\,$}}T{<}2$ through
much of this energy range. 

To determine whether this strong transmission is
occurring mainly by direct electron transmission
from molecule to molecule along the chain or mainly
through the silicon substrate the calculation was
repeated with all Hamiltonian matrix elements and
overlaps between atomic orbitals on different
molecules switched off so as to allow conduction
along the chain to proceed only via the substrate.
(Note that the parameters describing the ideal
leads and their coupling to the sample were kept
unchanged.)   The result is shown in
Fig.\ref{Fig_9}(b). Clearly throughout most of the
LUMO energy band the transmission via the substrate
channel (Fig.\ref{Fig_9}(b)) is much weaker than
the direct transmission through the molecular chain
itself. The transmission through the substrate is
strong only at a small number of narrow resonances.
Furthermore at the energies at which the strongest
of these resonances occur the transmission of the
fully coupled system (Fig.\ref{Fig_9}(a)) shows
{\em minima}; this means that the resonant states
of the substrate that give rise to the strongest
transmission peaks in Fig.\ref{Fig_9}(b) give rise
to resonant {\em  back-scattering} (as opposed to
contributing to resonant forward transmission) in
the fully coupled system where transport is
permitted through both the molecular chain and
substrate. Thus it is quite clear that in the
energy range of the LUMO band of the molecular
chain, transport from one end of the molecular
chain to the other is strongly dominated by direct
electron transmission from molecule to
molecule.\cite{idealleads}

The corresponding results for transmission from one end
of the molecular chain to the other in the energy range
of the molecular HOMO band are shown in
Fig.\ref{Fig_9}(c) for the fully coupled system, and in
Fig.\ref{Fig_9}(d) for the system with all Hamiltonian
matrix elements and overlaps between atomic orbitals on
different molecules switched off. The molecular chain is
still a fairly good one-dimensional conductor ($T(E)\sim
1$ in most of this energy range in Fig.\ref{Fig_9}(c)).
Transmission through the substrate (Fig.\ref{Fig_9}(d))
is again clearly less efficient than through the
molecular chain although the difference between the two
is much less stark for the HOMO energy range than for the
LUMO.

A simple measure of the overall relative importance of
direct transmission from molecule to molecule along the
molecular chain relative to that via the parallel path
through the substrate is the ratio $R$ of the average
transmission for the fully coupled system to the average
transmission with the molecule-to-molecule coupling
switched off. For the 8 styrene molecule chain $R \sim
30$ for the molecular LUMO band while for the molecular
HOMO band $R \sim 4$.

For ideal source and drain leads coupled to opposite ends
of the substrate instead of to the molecular benzene
rings at the ends of the styrene chain, the calculated
average transmission probabilities in the HOMO and LUMO
energy ranges were found to be similar in magnitude. Thus
the above difference in $R$ between the HOMO and LUMO
bands of the molecular chain is mainly due to weaker
hybridization of the substrate orbitals with the
molecular LUMO states and than with the molecular HOMO.

Since the above calculations indicate that states of the
LUMO band of the molecular chain hybridize much less
with the states of the silicon substrate than the states
of the molecular HOMO band do, it is reasonable to expect
it to be easier for STM measurements to detect a
distinctive signature of electronic states confined to
the molecular chain and extending over a significant
distance along the molecular chain in the LUMO energy
range than in the HOMO range. The results of the
calculations of vertical transport in the STM geometry
presented in Section
\ref{vertical} are consistent with this: In particular,
one of the theoretical STM current plots (curve e) in Fig.
\ref{Fig_8} shows an overall decline in the STM current
as one moves away from the center of the molecular chain
in either direction; the decline extends all the way to
the end of the molecular chain in either direction but is
especially strong on the styrene side of the chain. At
the bias voltage corresponding to this current plot, the
Fermi level of the STM tip is slightly above the lower
edge of the LUMO band of the molecular chain and a single
transmission resonance due to lowest electronic
eigenstate belonging to the LUMO band makes the dominant
contribution to the current. The electron probability
distribution in this eigenstate is peaked at the center
of the molecular chain and declines strongly towards both
ends of the molecular chain. This explains qualitatively
why the calculated STM current (curve e, Fig.
\ref{Fig_8}) is stronger at the center of the chain than
at its ends. (However, to predict the current profile
along the chain {\em quantitatively} knowledge of the
LUMO eigenfunction alone is not sufficient and a complete
transport calculation such as is described in Section
\ref{transport} is necessary.) By contrast, the
calculated STM current for the case where the HOMO band
of the molecular chain is being probed does not show such
pronounced trends across more than a single molecule,
even when the tungsten Fermi level is close to the edge
of the HOMO band, just as one might expect qualitatively
for a molecular band that mixes more freely with states
belonging to the substrate. 
    
\section{End Effects}
\label{ends}

The model geometries of the molecular chains studied
theoretically in the preceding sections were
constructed (see Section \ref{structure}) to be
representative  of regions of the chain  far from its
ends. This was done by constraining the average
spacing between the tops of the molecules  to be
similar to that observed experimentally near the
center of the molecular chain. If this constraint is
{\em not} imposed, molecules at the ends of the chain
tilt more strongly towards the  silicon substrate.
This is seen experimentally in Fig. \ref{Fig_2}(a) and
(b), where the apparent difference in height between
the end and central molecules of the chain at low
negative sample bias is more than an Angstrom.
However, the observed height contrast between the 
ends and center of the chain in Fig. \ref{Fig_2}(a)
and (b) decreases markedly as the magnitude of the
bias increases. Thus considerations other than the
geometrical height profile of the molecular chain
must also be taken into account to explain this
data.   

The gaps between the molecular HOMO and LUMO bands
of stacks of parallel-oriented aromatic molecules
should increase with increasing separation between
the molecules because the coupling between adjacent
molecules decreases as their separation
increases.\cite{Rochefort} This suggests the
following scenario as a simple explanation of the
bias-dependence of the STM profiles of the molecular
chains that we have observed: The  tilting apart of
the molecules near the ends of the chains causes the
local HOMO-LUMO band gap of the molecular chain to
increase towards the ends of the chain. In constant
current STM images at low bias where the Fermi level
of the tip  is in the molecular HOMO-LUMO band gap,
this increased band gap causes the molecular chain
to appear {\em lower} at its ends relative to its
center than the geometrical difference in height
alone would require. At larger bias, when the tip
Fermi level has fallen below the (local) top  of the
molecular HOMO band throughout the molecular chain 
and {\em all} of the molecules have ``turned on,"
the end molecules appear  less diminished. It
appears therefore that the variable HOMO-LUMO gap
scenario  could account qualitatively  for the
position and voltage dependent height variations
observed.  However, a more thorough analysis shows
this model to be incomplete; it does not take into
account three aspects of this system that we show
below to be crucial.   They are: (1) The molecules
bond to equally spaced Si dimers on the (001) silicon
surface. Thus although the  spacing of the tops of
the molecules in STM images varies along the chain
due to the varying tilt of the molecules from the
surface normal, the spacing between the bottoms of
the molecules is fixed and thus, as will be seen
below, even the most strongly tilted molecule at the
end of the chain couples strongly to the rest of the
chain. (2) The coupling of molecular HOMO and LUMO
orbitals to the substrate (to which vertical
transport is very sensitive) increases strongly
along the chain as its end is approached because of
the increasing tilt of the molecules towards the
substrate.\cite{Korn} (3) At low bias where the STM
tip Fermi level is in the molecular HOMO-LUMO gap the
transmission of electrons between the substrate and
tip is mediated by resonant states at the C-Si
interface rather than the molecular HOMO and LUMO
orbitals, as is discussed in Section
\ref{vertical}; i.e., a different transport mechanism
that is not included in the pure variable band gap
scenario outlined above operates in the low bias
regime.         
                
For simplicity, we restrict our attention to styrene
chains.   Our results for  the (vertical) STM
current over  a chain of 8 styrene molecules on a
hydrogen-terminated dimerized Si(100) substrate for
negative  sample bias are shown in Fig.
\ref{Fig_10}. The silicon substrate, tungsten STM tip
and ideal source and drain leads are  as in Section
\ref{vertical}. However, the structure of the
molecular chain is different: Here the final AM1
relaxation of the chain differed from that described
in  Section \ref{structure} in that the styrene
molecule at the left end of the chain was constrained
to stand upright on the substrate (with a small tilt
to the right) but the other molecules were not
constrained in this way. Thus the structure of this
model styrene chain near its left end resembles that
in the interior of long molecular chain (as for the
model structures considered in Section
\ref{vertical})  while its right end should be
similar to that of the real end of a molecular chain.
The currents shown in Fig. \ref{Fig_10} are
calculated for STM tip trajectories that pass
directly over the 4th carbon atom of the benzene
ring of each styrene molecule, i.e.,  the carbon
atom furthest from the carbon chain that anchors the
benzene ring to the silicon. [The locations of these
carbon atoms are shown by dotted vertical lines in
Fig. \ref{Fig_10}.]  This 4th carbon atom for the
molecule at the right end of the model styrene chain
is located 1.12 \AA closer to the Si substrate than
the highest carbon atom of the
chain,\cite{height,4th} a height difference similar
to the apparent height contrast between the end  and
interior molecules of the chain observed
experimentally at low bias; see
Fig.\ref{Fig_2}(a,b). The current plots L (H) in
Fig. \ref{Fig_10} are for the same low (high) values
of the negative sample bias as the plots L (H) in
Fig. \ref{Fig_6}; for L (H) the STM  tip Fermi level
is above (within) the molecular HOMO band as in Fig.
\ref{Fig_5}. However, in Fig. \ref{Fig_10} the
currents labelled H have been scaled up by a factor
of 5 for clarity while that in Fig.  \ref{Fig_6} was
scaled down by  a factor of 10.  The solid curves L
and H in Fig. \ref{Fig_10} show the  calculated STM
current for scans at constant tip height above the Si
substrate (the same height as for curves L and H in
Fig. \ref{Fig_6}).        The dashed curves,
however, show calculated currents for tip
trajectories  which follow the sloping height
profile of the relaxed molecular chain (thus  at
varying height from the underlying silicon
surface).\cite{heightoverC4} 

For the scans at constant height above the silicon
surface (the solid curves) the calculated  current
over the tilted molecule at the right end of the
chain is lower than that near the left end, by a
factor of more than 30 for the low bias case L and a
factor of $\sim 10$ for the higher bias case H. These
results agree qualitatively with our experimental
finding that the height of the molecules appears
lower  at the end of the chain and that the magnitude
of the lowering decreases as the magnitude of the
bias increases. However for the lower bias scan at a
constant vertical distance above the tops of the
molecules  (dashed curve L) the calculated current
does not decrease near the right end of the
molecular chain; in fact it is slightly {\em higher}
over the molecule that ends the chain than over the
rest of the chain. By contrast, for the simple
position-dependent band gap model outlined above to
apply in isolation the apparent height of the
molecules  at the end of the chain  {\em must} be
substantially {\em lower} than their geometrical
height. Thus our calculations do not support a
purely position-dependent band gap scenario. They
suggest instead   that at low negative substrate
bias in the constant current STM mode the
experimentally measured differences in the height of
the tip from styrene molecule to styrene molecule
along the chain should, to a first approximation, be
equal to the differences in the geometrical heights
of the tops of the respective molecules.  This
interpretation is further supported by the
approximate agreement between the apparent height
difference in experimental data in Fig.
\ref{Fig_2}(a,b) between the styrene molecules near
the center and end of the chain at low bias  and the
result of our calculation of the structure at the end
of the molecular chain: In each case the top of the
molecule at the end of the chain is lower than the
tops of the molecules far from the ends by somewhat
more than 1 \AA.

We explain the bias dependence of the apparent
height of the molecules at the end of the chain
relative to those far from the end as follows:
Because the molecules near the end of the chain tilt
more strongly towards the substrate than the other
molecules of the chain do, the molecular HOMO
orbitals of the end molecules (being mainly on the
benzene rings) couple much more strongly to the 
substrate than the HOMO orbitals of the other
molecules do. This stronger coupling leads to
stronger electron transmission from the substrate to
the STM tip through the end molecules of the chain
than through the other molecules at the higher
values of the negative substrate bias for which the
Fermi level of the tungsten tip falls below the top
of the molecular HOMO band so that transport via the
HOMO orbitals  dominates the
conduction.\cite{lowbiasres} Thus for the dashed
curve H in Fig. \ref{Fig_10} where the tip scans
along the chain at same vertical distance above each
molecule, the current grows quite significantly as
the end of the chain is approached. Consequently the
decrease in the apparent height of the molecules as
the end of the chain is approached in  {\em constant
current} STM mode is not as strong as it is at lower
bias where the molecular HOMO states do not mediate
the transport.\cite{lowbiasres} 

This mechanism is clarified further by the inset of
Fig.
\ref{Fig_10}: Here the Landauer transmission $T(E,V)$
between the substrate and STM tip is plotted as a
function of electron energy E for the same (higher)
value of the applied bias $V$ as for curves H of Fig.
\ref{Fig_10}. For each transmission curve plotted
the STM tip is located 3.497 \AA  above the 4th
carbon atom of the benzene ring of the nearest
molecule (as for the dashed curve H of Fig.
\ref{Fig_10}). For reference the arrow indicates the
HOMO energy of an ethyl-benzene molecule in free
space.    The solid line is the calculated
transmission with the tip above the molecule at the
right end of the molecular chain (with the other
molecules of the chain also present on the
substrate). The dashed line shows the transmission
with the tip above the same molecule (with the same
tilted geometry) but with the other molecules of the
chain removed from the system and replaced by
hydrogen atoms. The dot-dashed curve is the average
transmission for tip positions over the 5 molecules
closest to the left end of the molecular chain. The
nearly equally spaced  peaks of the dot-dashed curve
arise from states of the molecular HOMO band that is
formed due to the coupling between the HOMO orbitals
of the molecules making up the
chain.\cite{discrete}  The much greater strength of
the transmission for the tip above the molecule at
the right end of the chain (solid curve) is due to
the stronger coupling of that molecule's HOMO
orbital to the substrate, as discussed above. It
results in the stronger  current when the tip is
over the right end molecule than over the rest of
the chain, as is seen in the dashed curve H. The
stronger coupling of the right end molecule's HOMO
to the substrate also results in the strong
broadening of the the associated transmission peaks
(solid and dashed lines of the inset); the HOMO
transmission peaks (not shown) for isolated
molecules on the substrate with the upright
geometries of molecules far from the end of the
chain are much narrower, with widths similar to
those of the individual peaks of the dot-dashed
curve of the inset. 

Although, as we have shown, the enhanced coupling of
the molecular HOMO at the end of the chain to the
substrate  plays a crucial role in the bias
dependence of the STM profile of the end of the
chain, the coupling of the end molecule to the rest
of the molecular chain remains strong despite the
tilting of the end molecule towards the substrate.
This is evident from a comparison of the
transmission peaks for the end molecule with the
rest of the molecular chain present and absent from
the system: The coupling of the end molecule to the
rest of the chain results in the large splitting of
the transmission peak that occurs when the rest of
the molecular chain is present (solid curve of the
inset) and does not occur when the rest of the chain
is absent (dashed curve).\cite{decouple,spittingimp} 

Our results for positive substrate bias are shown in
Fig. \ref{Fig_11}. The chain geometry and the
meaning of the solid and dashed plots are as in the
negative bias case (Fig. \ref{Fig_10}). As in  Fig.
\ref{Fig_8},  for plots l, e and h the STM tip Fermi
level is below, slightly above and well above the
lower edge of the molecular chain's LUMO band (see
Fig. \ref{Fig_5}), but in Fig.
\ref{Fig_11} the currents l have been scaled down by
a factor of 10. The molecules at the end of the chain
are again predicted to appear lower in STM images than
those near the center, and this lowering becomes
less pronounced as the magnitude of the applied bias
voltage increases.\cite{novargap}  The mechanism of
the bias dependence is similar to that for negative
substrate bias: Because of the tilting of the
molecules at the end of the chain, their LUMO
orbitals couple more strongly to the substrate, and
this results in stronger and broader transmission
peaks (and an enhanced STM current) over the end
molecules when the applied bias is large enough for
the LUMO orbitals to participate in conduction.
However, for plots e and h of Fig.
\ref{Fig_11} the local current {\em maxima} occur
over or near the tops of the molecules near the end
of the chain while far from the end current {\em
minima} coincide with the molecular positions. 
Because of the long distance that the influence of
the physical (right) end of the molecular chain
extends along the chain in plot e, experimentally
distinguishing effects related to the presence of a
heterojunction from such end effects in empty-state
STM imaging will require  long molecular chains and
careful analysis.

\section{Conclusions}
\label{conclusions}
We have synthesized heterojunctions of lines of styrene
and methyl-styrene molecules on H-terminated (100)
silicon substrates and have studied these systems with
scanning tunneling microscopy as well as theoretically.
We have developed a tight binding theory based on the
extended H{\"u}ckel model (modified so as to provide a
good description  of the electronic structure of
tungsten and silicon crystals as well as of the
molecules)  that together with Green's function and
Landauer transport techniques is able to account for key
features of the experimental filled-state STM images of
these molecular systems (including both the
heterojunctions and the regions near the ends of the
molecular chains) and of their variation with applied
bias voltage. This model has also been used to explore
theoretically the lateral transport properties of short
chains of styrene molecules on silicon and it has been
found that electron transmission along such chains should
occur primarily from molecule to molecule rather than
through the parallel substrate channel, especially in the
energy range of the molecular LUMO band. The calculations
also suggest that phenomena due to electron wave
functions extending from molecule to molecule (such as
states localized around the heterojunction of a pair of
molecular wires) may be observable experimentally by STM
imaging.
\section*{Acknowledgments}
This research was supported by the Canadian Institute
for Advanced Research, NSERC, iCORE and the NRC. We
have benefited  from discussions with G. DiLabio and J.
Pitters and from the technical expertise of D.J.
Moffatt, and M. Cloutier.

%
%

%
%
\begin{figure*}
\caption{(Color on line). Constant current
filled-state STM imaging of a methyl-styrene/styrene
heterostructure on the H:Si(100) surface. {\bf a},
Growth of the first line segment of methyl-styrene
(8L exposure). The chemically reactive silicon
radical which initiated the line growth (not shown)
is now located at the end of the CH$_3$-styrene
segment (indicated by white arrow). This silicon
radical will serve as the initiation site for the
growth of the second line segment. {\bf b}, Following
a 78 L exposure of styrene, the methyl-styrene segment
in panel (a) has been extended to form a molecular
heterostructure. The white wedge indicates the
location of the heterointerface. At V$_s$ = -3.0 V,
the methyl-styrene line segment images higher than
the styrene segment. {\bf c}, V$_s$ = -2.4 V, the
methyl-styrene and styrene line segments image with
similar height. {\bf d}, V$_s$ = -1.8 V, the
methyl-styrene and styrene line segments reveal
different molecular contrast. Inset: Blue and red
arrows show locations of topographic cross-sections
presented in Figures \ref{Fig_2} (a) and (b),
respectively. Images were acquired using a constant
tunnel current of 60 pA. Image areas: 8.5
nm$\times$8.5 nm.}
\label{Fig_1}
\end{figure*}


\begin{figure*}
\caption{(Color on line). Topographic cross-sections
extracted from STM image data of the methyl-styrene/
styrene heterostructure presented in Figure 1. {\bf
a}, Topographic cross-sections taken along the
chemical attachment points between the
heterostructure and the underlying dimer row - along
direction of blue arrow in Figure \ref{Fig_1} (d).
{\bf b}, Topographic cross-sections taken to the
right of the trench separating the underlying dimer
rows - along direction of red arrow in Figure
\ref{Fig_1} (d). Cross-sections for a range of sample
biases (-3.0 V to -1.8 V) are shown in each panel.
Black arrows in panels (a) and (b) indicate the
position of the interfacial methyl-styrene molecule.}
\label{Fig_2}
\end{figure*}

%
%
%
\begin{figure*}
\caption{(Color on line). Model structure of a
heterojunction between methyl-styrene and styrene
chains (each of four molecules) on a
hydrogen-passivated silicon cluster representing a
dimerized H-terminated Si (001) surface. Carbon atoms
are black, silicon is blue, hydrogen is white. Inset:
Another view of the molecules and nearby substrate
atoms.}
\label{Fig_3}
\end{figure*}

%
%
\begin{figure*}
\caption{(Color on line).Geometry of a methyl-styrene
molecule and styrene molecule relaxed using
density functional theory on a small dimerized Si
(001) cluster. Notation as in Fig.\ref{Fig_3}. For
significance of the numbering see text.}
\label{Fig_4}
\end{figure*}

%
\begin{figure*}
\caption{Histogram of the silicon (a), styrene carbon
(b), methyl-styrene carbon (c) and hydrogen (d)
content of the electronic eigenstates of the
eight-molecule styrene/methyl-styrene chain on the
hydrogen passivated silicon cluster depicted in
Fig.\ref{Fig_3}. The arrows labelled H and L indicate
the positions of the tungsten Fermi level relative to
the molecular energy spectrum at the negative
substrate biases corresponding to current curves H and
L,L$'$ in Fig.\ref{Fig_6}. Arrows ll, l, e and h show
the tungsten Fermi level at the positive substrate
biases corresponding to current curves ll, l, e and h
in Fig.\ref{Fig_8}.}
\label{Fig_5}
\end{figure*}

%
\begin{figure*}
\caption{Calculated current $I$ (arbitrary units)
flowing between the tungsten STM tip and the
styrene/methyl-styrene molecular chain on silicon at
some negative substrate biases vs. STM tip position
along the chain. The current for curve H has been
scaled down by a factor 10. Dotted vertical lines
labelled m (s) indicate the locations of the highest
carbon atoms of the methyl-styrene (styrene)
molecules. The scans are at constant tip height,
2 {\AA} (3 {\AA}) higher than the highest carbon atom of
the chain for curves H and L (curve L$'$). L and L$'$
are at a lower bias, H at a higher bias. The tungsten
Fermi level at each bias is indicated by an arrow in
Fig.\ref{Fig_5}; see text.}
\label{Fig_6}
\end{figure*}

%
\begin{figure*}
\caption{Calculated current  between the tungsten STM
tip and the styrene/methyl-styrene molecular chain on
silicon vs. STM tip position scanned across the chain,
in the direction perpendicular to that of the scans
in Fig.\ref{Fig_6} and
\ref{Fig_8}.  Notation as in Fig.\ref{Fig_6}. The
scans are at constant tip height of 2 {\AA} higher than
the highest carbon atom of the molecular chain. (a)
Scan sL (mL) is over the third styrene
(methyl-styrene) molecule from the junction for the
same bias as scan L in Fig.\ref{Fig_6}. (b) Scan se
(me) is over the third styrene (methyl-styrene)
molecule from the junction for the same bias as scan e
in Fig.\ref{Fig_8}. Each scan passes over the highest
carbon atom of the respective styrene or
methyl-styrene molecule and crosses the locus of the
corresponding scan of Fig.\ref{Fig_6} or
\ref{Fig_8} at position = 0.}
\label{Fig_7}
\end{figure*}

%
\begin{figure*}
\caption{Calculated current  between the tungsten STM
tip and the styrene/methyl-styrene molecular chain on
silicon at some positive substrate biases vs. STM tip
position along the chain. Notation as in
Fig.\ref{Fig_6} but the scale is the same for all
curves. The scans are at constant tip height of 2 {\AA}
higher than the highest carbon atom of the molecular
chain. The magnitude of the bias voltage increases
from curve ll to l to e to h (see text). The tungsten
Fermi level at each bias is indicated by an arrow in
Fig.\ref{Fig_5}. }
\label{Fig_8}
\end{figure*}

\begin{figure*}
\caption{Calculated Landauer transmission
probability
$T(E)$ (per spin channel) through  a molecular
chain of 8 styrene molecules on a
hydrogen-terminated silicon cluster  in the
energy range of the molecular LUMO band (a) and
of the molecular HOMO band (c). Transmission in
the same energy ranges with all Hamiltonian
matrix elements and overlaps between orbitals of
different molecules switched off (so that
transport between the terminal molecules must
proceed via the substrate) is shown in plots (b)
and (d), respectively.}
\label{Fig_9}
\end{figure*}

\begin{figure*}
\caption{Calculated current $I$ (arbitrary units)
flowing between the tungsten STM tip and a styrene
molecular chain on silicon at some negative substrate
biases vs. STM tip position along the chain projected
onto an axis parallel to the Si dimer rows of the
substrate. The tungsten Fermi level at each bias (L
or H) is indicated by an arrow in Fig.\ref{Fig_5}.
The styrene molecule at the left end of the
chain has been constrained to stand
approximately upright in a geometry
representative of the region near the center of
a molecular chain while the other molecules of
the chain have been allowed to relax so that the
molecular geometries at right end of the chain
are representative of the physical end of a
molecular chain on silicon. (See text).   The
currents for curves H has been scaled up by a
factor 5. Dotted vertical lines labelled s
indicate the lateral locations of the carbon
atoms of the styrene molecules that are furthest
from the carbon chains anchoring the molecules
to the silicon. The dashed curves are for scans
in which the tip passes  over each styrene
molecule at an equal vertical distance (3.497
\AA)  above this carbon atom. The solid curves
are for scans in which the tip is kept at a
constant height above the silicon substrate.
Inset: The solid curve shows the Landauer
transmission probability between tip and
substrate vs. electron energy in eV with the STM
tip 3.497 \AA above the styrene molecule at the
right end of the chain for the same applied bias
as curves H. Dashed curve: Same except that all
molecules other than the molecule at the right
end of the chain are replaced with H atoms
passivating the Si substrate. Dot-dashed curve:
Landauer transmission probability averaged over
tip positions 3.497 \AA above the five styrene
molecules nearest the left end of the molecular 
chain.}
\label{Fig_10}
\end{figure*}

\begin{figure*}
\caption{Calculated current $I$ flowing between the
tungsten STM tip and a styrene molecular chain on
silicon at some positive substrate biases vs. STM tip
position. The tungsten Fermi level at each bias (l, e
and h) is indicated by an arrow in Fig.\ref{Fig_5}. 
Notation as in Fig.
\ref{Fig_10}. }
\label{Fig_11}
\end{figure*}


\begin{thebibliography}{100}
\bibitem{Bumm}L. A. Bumm, J. J. Arnold, M. T. Cygan, T.
D. Dunbar, T. P. Burgin, L. Jones II, D. L. Allara, J. M.
Tour, and P. S. Weiss, Science {\bf 271}, 1705 (1996).
\bibitem{Reed1} M.~A.~Reed, C.~Zhou, C.~J.~Muller,
T.~P.~Burgin, and J.~M.~Tour, Science {\bf 278}, 252
(1997).
\bibitem{Datta} S.~Datta, W.~Tian, S.~Hong,
R.~Reifenberger, J.~I.~Henderson, C.~P.~Kubiak, Phys.
Rev.  Lett. {\bf 79}, 2530 (1997).
\bibitem{Reedsw} J.~Chen, M.~A.~Reed, A.~M.~Rawlett and
J.~M.~Tour, Science {\bf 286}, 1550 (1999).
\bibitem{Collier} C.~P.~Collier, G.~Mattersteig,
E.~W.~Wong, Y.~Luo, K.~Beverly, J.~Sampaio, F.~M.~Raymo,
J.~F.~Stoddart and J.~R.~Heath, Science {\bf 289}, 1172
(2000).
\bibitem{Donhauser} Z.~J.~Donhauser, B.~A.~Mantooth,
K.~F.~Kelly, L.~A.~Bumm, J.~D.~Monnell, J.~J.~Stapelton,
D.~W.~Prince Jr., A.~M.~Rawlett, D.~L.~Allara,
J.~M.~Tour and P.~S.~Weiss, Science {\bf 292}, 2303
(2001).
\bibitem{Wolkow} For a review see R. A. Wolkow, Annu.
Rev. Phys. Chem. {\bf 50}, 413 (1999).
\bibitem{Lopinski}G. P. Lopinski, D.D.M. Wayner, R.A.
Wolkow, Nature {\bf 406}, 48 (2000).
\bibitem{Cho}J.-H. Cho, D.-H. Oh, and L. Kleinman,Phys.
Rev. B {\bf 65}, 081310(R) (2002).
\bibitem{Hofer}W. A. Hofer, A. J. Fisher, G. P.
Lopinski, R. A. Wolkow, Chem. Phys. Lett. {\bf 365},129
(2002).
\bibitem{Hersam}N. P. Guisinger, M. E. Greene, R. Basu,
A. S. Baluch, and  M. C. Hersam,  Nano Letters {\bf 4},
55 (2004).
\bibitem{Tong}X. Tong, G. A. DiLabio, R. A. Wolkow, Nano
Letters {\bf 4}, 979 (2004).
\bibitem{Rakshit}T. Rakshit, G.-C. Liang, A. W. Ghosh,
and S. Datta, Nano Letters {\bf 4}, 1803 (2004).
\bibitem{Wang}Y. Wang and G. S. Huyang, Appl. Phys Lett.
{\bf 86}, 023108 (2005).
\bibitem{Rochefort}A. Rochefort, R. Martel and P.
Avouris, Nano Letters {\bf 2}, 877 (2002).
\bibitem{Yaliraki}S. N. Yaliraki and M. A. Ratner, J.
Chem. Phys. {\bf 109}, 5036 (1998).
\bibitem{Magoga}M. Magoga and C. Joachim, Phys. Rev.
B{\bf 59}, 16011 (1999).
\bibitem{Lang}N. D. Lang and P. Avouris, Phys. Rev. B{\bf
62}, 7325 (2000).
\bibitem{Kushmerick1}J. G. Kushmerick, J. Naciri, J. C.
Yang, and R. Shashidhar, Nano Lett. 3, 897 (2003).
\bibitem{Liu1}R. Liu, S.-H. Ke, H. U. Baranger, and W.
Yang, J. Chem. Phys. {\bf 122}, 044703 (2005).
\bibitem{Boland}J. J. Boland, Surf. Sci. {\bf 261}, 17
(1992).
\bibitem{submitted}Submitted.  Also G.A.
DiLabio, $et$ $al.$ J. Am. Chem. Soc. {\bf 126}, 
16048 (2004) 
for examples of the
self-directed growth of linear alkane molecules.
\bibitem{Other}Other methyl-styrene/styrene
heterostructures have been studied (not shown) which
still possess the terminal silicon radical. They image
with similar characteristics but show locally perturbed
electronic structure at low magnitude bias resulting
from interaction with the terminal silicon
radical.\cite{Nature} Empty-state images (not shown) were
successfully acquired for a single heterostructure, and
will not be discussed.
\bibitem{Nature}P.G. Piva, G.A. DiLabio, J.L. Pitters,
J. Zikovsky, M. Rezeq, S. Dogel,  W.A. Hofer, R.A.
Wolkow, Nature {\bf 435}, 658 (2005).
\bibitem{reversebonding} The conformation of the styrene
and methyl-styrene molecules  depicted in
Fig.\ref{Fig_3} and studied theoretically in this
article is based on the bonding geometry of chains of
styrene molecules to the H-terminated Si (100) surface
adopted in a number of previous  studies of those
systems.\cite{Lopinski,Hofer,Wang} However,  a subtly
different structure for  the styrene chains on silicon
may also be possible.  In this alternate
structure,\cite{Cho,DiLabio} the carbon atom adjacent to
the carbon atom that bonds to the silicon is located not
above the trench  between the silicon dimer rows as in
Fig.\ref{Fig_3} but above the silicon dimer to which  the
molecule bonds.\cite{Cho} However the benzene ring of the
molecule is still located over the  trench between the
dimer rows. Which geometry is realized in practice
remains unclear: Density functional calculations are not
reliable enough to decide unequivocally which geometry
has the lower energy, and it is not known whether either
structure is favored over the other by growth kinetics of
the molecular rows. Comparison of STM images of the 
styrene/methyl-styrene heterojunctions with the results
of our modeling also does not unambiguously favor one of
these geometries over the other.     
\bibitem{DiLabio}G.A. DiLabio, private communication.    
\bibitem{Gaussian} The Gaussian98 package with the B3LYP
density functional and 6-31G(d) basis set were used.
\bibitem{AM1}The AM1 relaxations were carried out using
the Gaussian98 package.
\bibitem{just} The identification between the spacing of
the molecules as measured in the experimental STM images
and the lateral spacing of the top carbon atoms of the
molecules is justified by our calculations of the STM
current maps at low bias as discussed in Section
\ref{negative}.
\bibitem{Gausstip} The Gaussian98 package with the B3PW91
density functional and Lanl2DZ basis set was used.
\bibitem{metalcontacttheory1} E. G. Emberly, G.
Kirczenow, Phys. Rev. B {\bf 58}, 10911 (1998);
S.~N.~Yaliraki, M.~Kemp, M.~A.~Ratner, J. Am. Chem.
Soc. {\bf 121}, 3428 (1999); V.~Mujica, A.~E.~Roitberg,
M.~Ratner, J. Chem. Phys. {\bf 112}, 6834 (2000); L. E.
Hall, J. R. Reimers, N. S. Hush and K. Silverbrook, J.
Chem. Phys. {\bf 112}, 1510 (2000); M.~{Di Ventra},
S.~T.~Pantelides, and N.~D.~Lang, Phys. Rev. Lett. {\bf
84}, 979 (2000); P.~S.~Damle, A.~W.~Ghosh and S.~Datta,
Phys. Rev. B {\bf 64}, 201403 (R) (2001); J.~Taylor, H.~Guo,
J.~Wang, Phys. Rev. B {\bf 63}, 121104 (R) (2001);
P.~E.~Kornilovitch and A.~M.~Bratkovsky, Phys. Rev. B
{\bf 64}, 195413 (2001).
\bibitem{Emberly} E.~G.~Emberly and G.~Kirczenow, Phys.
Rev. Lett. {\bf 87}, 269701 (2001), Phys. Rev. B {\bf
64}, 235412 (2001).
\bibitem{metalcontacttheory2}M.~Paulsson, S.~Stafstrom,
Phys. Rev. B {\bf 64}, 35416 (2001); J. J. Palacios, A.
J. P\'{e}rez-Jim\'{e}nez, E. Louis, J. A. Verg\'{e}s,
Phys. Rev. B 64, 115411 (2001); M.~H.~Hettler,
H.~Schoeller, W.~Wenzel, Europhys. Lett. {\bf 57}, 571
(2002); J. Taylor, M. Brandbyge, K. Stokbro, Phys. Rev.
Lett. {\bf 89}, 138301 (2002); J. Heurich, J. C. Cuevas,
W. Wenzel, G. Sch\"{o}n, Phys. Rev. Lett. {\bf 88},
256803 (2002); R. Gutierrez, F. Grossmann, R. Schmidt, 
Chemphyschem {\bf 3}, 650 (2002).
\bibitem{Kushmeric}J.~G.~Kushmerick, D.~B.~Holt,
J.~C.~Yang, J.~Naciri, M.~H.~Moore, R.~Shashidhar, Phys.
Rev. Lett {\bf 89}, 086802 (2002).
\bibitem{metalcontacttheory3} E. G. Emberly and G.
Kirczenow, Phys. Rev. Lett. {\bf 91}, 188301 (2003);
Y. Xue and M.A. Ratner, Phys. Rev. B {\bf 69}, 085403
(2004), Phys. Rev. B  {\bf 70}, 081404 (2004); S.H. Ke,
H.U. Baranger, W.T. Yang {\bf 70}, 085410 (2004).  
\bibitem{Guo}C.C. Kaun, H. Guo, Nano Letters {\bf 3},
1521 (2003).
\bibitem{Lee}T. Lee, W. Wang, J. Klemic, J. Zhang, J. Su,
and M. A. Reed, J. Phys. Chem. B {\bf 108}, 8742 (2004).
\bibitem{DFTtouble} These difficulties may be related in
part to the fact that while Kohn-Sham theory is
appropriate for calculating total energies and relaxed
molecular geometries, the single-electron eigenenergies
and wave functions that appear in it do not have a
rigorous physical meaning.\cite{KohnSham} While they are
often used successfully to approximate the corresponding
properties of electronic quasi-particles  that play a
crucial role in transport, this approximation need not
be accurate in general.   
\bibitem{KohnSham} W. Kohn and L.J. Sham, Phys. Rev. {\bf
140}, A1133 (1965).
\bibitem{fit}Seven independent fitting parameters
$K_{ij}$ with values falling in the range between 1.0 and
2.0 were found to suffice for this purpose.
\bibitem{pap}The tungsten Hamiltonian matrix elements
were fitted to the band structure of bulk bcc tungsten
given by D.A. Papaconstantopoulos, in {\em Handbook of
the Band Structure of Elemental Solids}, Plenum Press,
New York, 1986. The Fermi level of the model tungsten STM
tip at zero bias was set 0.45 eV below the calculated HOMO
level of the 15 atom tungsten cluster used to model the
tip to improve the match between features of the density
of states of the model tip and those of bulk tungsten
around the Fermi level.  
\bibitem{shift}E. G. Emberly, G. Kirczenow, Chem. Phys.
{\bf 281}, 311 (2002), Appendix A. 
\bibitem{drop}In reality a part of the potential drop
should occur within the molecule  and it has been
proposed that this may result in negative differential 
resistance (NDR) in STM experiments involving molecules
on silicon.\cite{Rakshit,Hersam} Since NDR has not been
observed experimentally in  the styrene and
methyl-styrene molecular chains on  silicon in this and
previous\cite{Lopinski} work, it seems reasonable to
neglect the bias-induced potential  drop across the
molecules, as a first approximation, for the bias
voltages considered here.     
\bibitem{book}For a review see S. Datta, {\em Electronic
Transport in Mesoscopic Systems}, Cambridge University
Press: Cambridge, 1995.
\bibitem{TEV}The applied bias shifts the energy levels of the
STM tip relative to those of the Si substrate and molecules.
Since we use a realistic tight-binding model to describe the
tungsten STM tip, this relative shift of the energy levels
results in the transmission coefficient
$T(E,V)$ depending on the value of the applied bias $V$ as
well as on the electronic energy $E$.
\bibitem{temp}The temperature $T$ in the Fermi functions
was set to 0 in the calculations reported here.
\bibitem{Buttikerprobes}M. B{\"u}ttiker, Phys. Rev.
B{\bf 33}, 3020 (1986).
\bibitem{orthog1}E. Emberly and G. Kirczenow, Phys.
Rev. Lett. {\bf 81}, 5205 (1998).
\bibitem{orthog2}E. G. Emberly and 
G. Kirczenow, J. Phys.: Condens. Matter
{\bf 11}, 6911 (1999).
\bibitem{traj} The trajectory of the tip is  above
approximately the centerline of the ``trench'' between
the two rows of Si dimers in Fig.\ref{Fig_3}.
\bibitem{Korn} An analogous effect has been proposed for
aromatic molecules on metal substrates by P. E.
Kornilovitch and A. M. Bratkovsky, Phys. Rev. B {\bf 64},
195413 (2001). 
\bibitem{heightvscurrent} Since the STM current at
constant bias is a decreasing function of tip
height, regions of higher current at constant tip
height coincide with regions of greater tip height
at constant current. 
\bibitem{nonode}The current node predicted to occur
at the center of each molecule at low bias by density
functional theory is absent in both the present
theoretical results and the corresponding
experimental data.
\bibitem{strongcoupl}The fact that curve L of
Fig.\ref{Fig_6} for which the tip is only 2 {\AA}
higher than the highest carbon atom of the molecular
chain resembles the experimental data at low bias
closely (whereas curve L$'$ for which the tip is 1
{\AA} higher than than for curve L does not)
underscores the importance of modeling the present
system without treating the electronic Hamiltonian
matrix elements and overlaps between the tip and
molecules as weak perturbations, as is  frequently
done in theories of STM imaging. The present
theoretical approach avoids making such a weak
coupling approximation as is explained at the start
of Section \ref{transport}.
\bibitem{othertips} In an alternate model of the STM
tip in which the tungsten atom  at the end of the tip
is replaced by a silicon atom, the calculated STM
current was also found to exhibit  a decreasing
contrast between the  methyl-styrene and styrene 
parts of the structure as the height of the tip above
the Si substrate or the magnitude of the tip  bias
was reduced. However, for a Si-terminated tip the
calculated current over a methyl-styrene molecule was
always  higher than that over a styrene molecule (for
equal tip heights above the silicon substrate),  in
contrast to curve L of Fig.\ref{Fig_6} for an
all-tungsten STM tip and to the low bias data in Fig.
\ref{Fig_2} (a).
\bibitem{moreaboutdips} The effect of the wave function nodes
of the molecular LUMO orbitals on the current profiles in
Fig. \ref{Fig_8}  (where the current corrugation is {\em
reversed} in plots e and h relative to that in l and ll) is
clearly more pronounced than that  of the molecular HOMO
orbitals on the current profiles in Fig. \ref{Fig_6}  (where
the corrugation in plots H is only weakened relative to that
in L). We attribute this difference in part to the well
known different behavior of the barrier for tunneling between
the molecules and an STM tip for positive and negative
substrate bias:\cite{GRIFFITH} When imaging filled sample
states, tunneling electrons from the sample with energies
just above the tip Fermi-level experience the largest tunnel
barrier. The current contribution from higher lying filled
sample states therefore tends to dominate the total tunnel
current.  In empty state imaging, the situation is reversed.
Tunneling electrons originating from the tip Fermi-level
experience the lowest tunnel barrier, dominate the tunneling
current, and offer a more sensitive probe of empty sample
states at that energy. Calculated current profiles are
therefore expected to be more sensitive to the details of
the sample wave functions at the tip Fermi-energy under
positive sample bias. However, in a quantitative analysis of
the effect, details of the tip electronic structure must be
included as well.
\bibitem{GRIFFITH}See J. E. Griffith and G. P. Kochanski,
Ann. Rev. Mat. Sci. {\bf 20}, 219 (1990).
\bibitem{idealleads}We note that the overall strength of the
calculated transmission is not sensitive to the precise
values of the parameters describing the ideal leads (although
the {\em fine} details of the transmission plots depend on
these parameters, as in other mesoscopic systems). This is
because a rather large number (18) of ideal leads is coupled
to each end of the molecular chain whereas the Landauer
transmission of the molecular chain is much smaller (less
than about 2 in Fig.\ref{Fig_9}, ignoring spin). Thus the
molecular chain and its Si substrate (rather than the ideal
leads connected to it) constitute the main transport
bottleneck in this system, and therefore the precise values
of the lead parameters are not very important so long as the
choices made for them are reasonable, for example, the band
edges for the ideal leads should not be very close to the
energy range being studied.   
\bibitem{height} It should be noted that the quoted
height of the carbon atom of the end molecule relative
to those of molecules far from the end applies to
long molecular chains.  Our calculations for short
chains yield structures in which the end molecules
tilt considerably less towards the silicon substrate.
\bibitem{4th} The 4th C atom of the benzene ring of
the end molecule is located
$\sim$0.07 \AA {\em lower} than one of its neighboring
C atoms on the same benzene ring but the STM current
above that neighboring atom is considerably lower,
which we attribute to a destructive quantum
interference effect at that site of the benzene ring
that has recently been pointed out:\cite{Cardamone}
An electron entering the benzene ring at its 1st
carbon atom has two paths that it can take around the
ring to its opposite end. Both paths to the 4th C atom
have the same length so the interference at that C
site is always constructive. But the two paths to the
carbon atoms next to the 4th C atom have different
lengths so there the interference can be destructive.
Thus this mechanism generally favors the 4th C atom.
\bibitem{heightoverC4}Currents were  evaluated in
this case for a vertical tip distance of 3.497 {\AA}
above the 4th carbon  atom of each benzene ring. 
\bibitem{Cardamone}D. M. Cardamone,  C. A. Stafford
and  S. Mazumdar, (2005) cond-mat/0503540.    
\bibitem{lowbiasres} As noted in Section
\ref{negative} the conduction mechanism at lower bias
is different, mediated by resonant states localized
primarily in the Si substrate and adjacent carbon
atoms rather than on the molecular HOMO orbitals,
thus the tilting of the molecules has a smaller effect
on electron transmission between substrate and tip
(at constant tip height above the molecule) in the low
negative substrate bias regime.
\bibitem{discrete}These band states are discrete
because of the finite length of the molecular chain;
for an infinite chain they would form a continuum.
\bibitem{decouple} The splitting was also found to be
absent when the remainder of the molecular chain is
present on the substrate but its direct coupling to
the end molecule is switched off by setting the
relevant matrix elements of the Hamiltonian and
overlap matrix to zero in calculating the
transmission.
\bibitem{spittingimp}The splitting affects the
magnitude of the STM current when the tip is over
the end of the molecular chain since the tip Fermi
level for the value of the applied bias in the inset
of Fig.
\ref{Fig_10} is close to the energy indicated by the
arrow, i.e., in the region between the two split
peaks of the solid transmission curve. 
\bibitem{novargap}Even at low positive substrate bias
the apparent height difference between the end and
center of the chain is somewhat {\em smaller} than
the geometrical height difference, again contrary to
the prediction of the simple variable band gap model.

   

\end{thebibliography}
\end{document}